\PassOptionsToPackage{table}{xcolor}
\documentclass[]{style}

\usepackage{amsmath,amssymb,amsthm}
\usepackage{booktabs,multirow,array,colortbl,tabularx,makecell}
\usepackage{xspace}
\usepackage{needspace}
\usepackage{wrapfig}
\usepackage{float}
\usepackage{enumitem}
\usepackage{tcolorbox}
\tcbuselibrary{skins,breakable}
\usepackage{titletoc}

\providecolor{pevIndigo}{HTML}{4440C6}
\providecolor{pevViolet}{HTML}{6B67EE}
\providecolor{pevPale}{HTML}{D4D3FA}
\providecolor{pevMist}{HTML}{F0F0F8}
\providecolor{pevWash}{HTML}{F9F9FE}
\providecolor{pevInk}{HTML}{1A1A2E}
\providecolor{pevTint}{HTML}{EBEBFE}   
\definecolor{pevRow}{HTML}{E4E3FB}     
\definecolor{pevHead}{HTML}{EEEDFC}    
\definecolor{pevStripe}{HTML}{F7F7FC}  

\AtBeginDocument{\setlength{\parfillskip}{0pt plus 0.7\textwidth}}

\titlespacing*{\paragraph}{0pt}{1.2ex plus 0.3ex minus 0.2ex}{0.6em}

\newcommand{\wrapguard}[1]{\Needspace*{#1\baselineskip}}
\AtBeginDocument{\newlength{\bodyparskip}\setlength{\bodyparskip}{\parskip}}

\newcommand{\tablestyle}{\small\setlength{\aboverulesep}{0pt}\setlength{\belowrulesep}{0pt}%
  \renewcommand{\arraystretch}{1.3}\rowcolors{2}{white}{pevStripe}}

\newcolumntype{O}{>{\columncolor{pevRow}}c}
\newcommand{\headrow}{\rowcolor{pevHead}}
\newcommand{\hd}[1]{{\leavevmode\sffamily\bfseries\footnotesize\color{pevInk}#1}}
\newcommand{\hdr}[1]{\multicolumn{1}{r}{\hd{#1}}}

\newcommand{\hdtwo}[1]{\hd{\shortstack{\rule{0pt}{2.1ex}#1}}}
\newcommand{\hdgroup}[2]{\multicolumn{#1}{c}{\hd{#2}}}

\newcolumntype{Y}{>{\raggedright\arraybackslash}X}
\newcolumntype{H}{>{\sffamily\bfseries\footnotesize\color{pevInk}}l}
\newcolumntype{C}[1]{>{\centering\arraybackslash}p{#1}}
\newcolumntype{R}[1]{>{\raggedleft\arraybackslash}p{#1}}
\newcommand{\ceilrow}[1]{\textit{\textcolor{pevInk!62}{#1}}}

\newif\ifnotes     \notestrue       
\newif\iffourpage  \fourpagefalse   
\newcommand{\longonly}[1]{\iffourpage\else #1\fi}

\newcommand{\note}[1]{\ifnotes\par\noindent{\color{red}{\bfseries\fbox{Outline}} {\itshape #1}}\par\fi}

\newcommand{\defnum}[2]{\expandafter\gdef\csname numv:#1\endcsname{#2}}
\DeclareRobustCommand{\num}[1]{\ifcsname numv:#1\endcsname\csname numv:#1\endcsname
  \else\textbf{??}\PackageWarning{numbers}{Missing number #1}\fi}

\DeclareRobustCommand{\Pev}{Pev\xspace}
\DeclareRobustCommand{\STOP}{\textsf{STOP}\xspace}
\newcommand{\pchoice}{p_{\mathrm{choice}}}
\newcommand{\pimp}{p_{\mathrm{improve}}}
\newcommand{\calA}{\mathcal{A}}
\newcommand{\calO}{\mathcal{O}}

\newcommand{\E}{\mathbb{E}}
\DeclareMathOperator{\softmax}{softmax}
\DeclareMathOperator*{\argmax}{arg\,max}
\newcommand{\Mode}[1]{\textcolor{pevViolet}{\texttt{#1}}}            
\newcommand{\HL}[1]{\colorbox{pevPale!55}{$\displaystyle #1$}}       
\newcommand{\best}[1]{\textbf{#1}}
\newcommand{\ours}{\rowcolor{pevRow}}

\newcommand{\up}{\,{\color{pevInk!55}$\uparrow$}}
\newcommand{\dn}{\,{\color{pevInk!55}$\downarrow$}}

\theoremstyle{definition}

\newtheorem{proposition}{Proposition}[section]
\theoremstyle{remark}

\tcbset{pevthm/.style={enhanced,breakable,colback=pevTint,colframe=pevViolet,boxrule=0pt,
  leftrule=0pt,arc=0pt,left=8pt,right=8pt,top=5pt,bottom=5pt,before skip=6pt,after skip=6pt}}
\tcolorboxenvironment{definition}{pevthm}
\tcolorboxenvironment{proposition}{pevthm}
\tcolorboxenvironment{remark}{pevthm}
\crefname{definition}{Definition}{Definitions}
\crefname{section}{Section}{Sections}\Crefname{section}{Section}{Sections}
\crefname{appendix}{Appendix}{Appendices}\Crefname{appendix}{Appendix}{Appendices}
\crefname{figure}{Figure}{Figures}\Crefname{figure}{Figure}{Figures}
\crefname{table}{Table}{Tables}\Crefname{table}{Table}{Tables}
\crefname{proposition}{Proposition}{Propositions}
\crefname{remark}{Remark}{Remarks}

\newtcolorbox{algobox}[1]{enhanced,breakable,colback=white,colframe=pevViolet,boxrule=0.8pt,
  arc=3pt,fonttitle=\sffamily\bfseries\small,coltitle=white,colbacktitle=pevViolet,title={#1},
  left=6pt,right=6pt,top=3pt,bottom=3pt,before skip=5pt,after skip=5pt,fontupper=\small}

\notesfalse
\fourpagefalse
\iffourpage
  \titlespacing*{\section}{0pt}{2.4ex plus 0.5ex minus 0.2ex}{1ex plus 0.1ex}
  \titlespacing*{\subsection}{0pt}{1.8ex plus 0.4ex minus 0.2ex}{0.8ex plus 0.1ex}\fi

\defnum{disc/panels}{45}  \defnum{disc/fam}{39}  \defnum{disc/cand}{33{,}463}
\defnum{disc/orc/hit}{0.949}   \defnum{disc/orc/enr}{18.94}  \defnum{disc/orc/auc}{1.000}  \defnum{disc/orc/wells}{1.0}
\defnum{disc/pev/hit}{0.424}   \defnum{disc/pev/enr}{7.19}    \defnum{disc/pev/auc}{0.729}  \defnum{disc/pev/wells}{6.1}
\defnum{disc/zs/hit}{0.231}    \defnum{disc/zs/enr}{2.80}     \defnum{disc/zs/auc}{0.545}   \defnum{disc/zs/wells}{48.3}
\defnum{disc/e1v/hit}{0.167}   \defnum{disc/e1v/enr}{2.16}    \defnum{disc/e1v/auc}{0.540}  \defnum{disc/e1v/wells}{53.9}
\defnum{disc/rnd/hit}{0.128}   \defnum{disc/rnd/enr}{1.02}    \defnum{disc/rnd/auc}{0.497}  \defnum{disc/rnd/wells}{27.8}
\defnum{disc/pev/hitpct}{42}   \defnum{disc/rnd/hitpct}{13}
\defnum{disc/hit_x_rnd}{3.3}   \defnum{disc/hit_x_zs}{1.8}   \defnum{disc/enr_x_zs}{2.6}
\defnum{disc/hit_x_e1v}{2.5}
\defnum{disc/wells_x_rnd}{4.5} \defnum{disc/wells_x_zs}{7.9}
\defnum{disc/pev/found}{4.2}  \defnum{disc/rnd/found}{1.3}  \defnum{disc/zs/found}{2.3}
\defnum{disc/m_rnd/hit}{0.334} \defnum{disc/m_rnd/hit_lo}{0.227} \defnum{disc/m_rnd/hit_hi}{0.444}
\defnum{disc/m_zs/hit}{0.218}  \defnum{disc/m_zs/hit_lo}{0.099}  \defnum{disc/m_zs/hit_hi}{0.342}
\defnum{disc/m_zs/auc}{0.209}  \defnum{disc/m_zs/auc_lo}{0.119}  \defnum{disc/m_zs/auc_hi}{0.297}
\defnum{disc/m_e1v/hit}{0.285} \defnum{disc/m_e1v/hit_lo}{0.162} \defnum{disc/m_e1v/hit_hi}{0.408}

\defnum{cls/dom/fam}{27}  \defnum{cls/dom/panels}{27}
\defnum{cls/dom/orc/hit}{0.941} \defnum{cls/dom/orc/enr}{25.47} \defnum{cls/dom/orc/auc}{1.000}
\defnum{cls/dom/pev/hit}{0.522} \defnum{cls/dom/pev/enr}{10.96} \defnum{cls/dom/pev/auc}{0.828}
\defnum{cls/dom/zs/hit}{0.219}  \defnum{cls/dom/zs/enr}{3.73}   \defnum{cls/dom/zs/auc}{0.549}
\defnum{cls/dom/e1v/hit}{0.141} \defnum{cls/dom/e1v/enr}{2.65}  \defnum{cls/dom/e1v/auc}{0.531}
\defnum{cls/dom/rnd/hit}{0.080} \defnum{cls/dom/rnd/enr}{1.05}  \defnum{cls/dom/rnd/auc}{0.498}
\defnum{cls/pep/fam}{7}   \defnum{cls/pep/panels}{13}
\defnum{cls/pep/orc/hit}{0.946} \defnum{cls/pep/orc/enr}{9.27}  \defnum{cls/pep/orc/auc}{1.000}
\defnum{cls/pep/pev/hit}{0.246} \defnum{cls/pep/pev/enr}{1.32}  \defnum{cls/pep/pev/auc}{0.572}
\defnum{cls/pep/zs/hit}{0.215}  \defnum{cls/pep/zs/enr}{0.87}   \defnum{cls/pep/zs/auc}{0.506}
\defnum{cls/pep/e1v/hit}{0.177} \defnum{cls/pep/e1v/enr}{1.24}  \defnum{cls/pep/e1v/auc}{0.535}
\defnum{cls/pep/rnd/hit}{0.199} \defnum{cls/pep/rnd/enr}{1.00}  \defnum{cls/pep/rnd/auc}{0.495}
\defnum{cls/ab/fam}{5}    \defnum{cls/ab/panels}{5}
\defnum{cls/ab/orc/hit}{1.000}  \defnum{cls/ab/orc/enr}{8.79}   \defnum{cls/ab/orc/auc}{1.000}
\defnum{cls/ab/pev/hit}{0.360}  \defnum{cls/ab/pev/enr}{2.16}   \defnum{cls/ab/pev/auc}{0.604}
\defnum{cls/ab/zs/hit}{0.340}   \defnum{cls/ab/zs/enr}{2.81}    \defnum{cls/ab/zs/auc}{0.623}
\defnum{cls/ab/e1v/hit}{0.280}  \defnum{cls/ab/e1v/enr}{1.90}  \defnum{cls/ab/e1v/auc}{0.604}
\defnum{cls/ab/rnd/hit}{0.202}  \defnum{cls/ab/rnd/enr}{0.93}   \defnum{cls/ab/rnd/auc}{0.495}
\defnum{cls/dom/hit_x_zs}{2.4}  \defnum{cls/dom/enr_x_zs}{2.9}  \defnum{cls/dom/hit_x_rnd}{6.5}
\defnum{cls/dom/m_rnd/hit}{0.442} \defnum{cls/dom/m_rnd/hit_lo}{0.312} \defnum{cls/dom/m_rnd/hit_hi}{0.578}
\defnum{cls/pep/m_rnd/auc}{0.091} \defnum{cls/pep/m_rnd/auc_lo}{0.028} \defnum{cls/pep/m_rnd/auc_hi}{0.164}
\defnum{cls/ab/m_rnd/hit}{0.158}  \defnum{cls/ab/m_rnd/hit_lo}{0.019}  \defnum{cls/ab/m_rnd/hit_hi}{0.261}

\defnum{obj/bind/panels}{21}  \defnum{obj/bind/fam}{15}
\defnum{obj/stab/panels}{24}  \defnum{obj/stab/fam}{24}
\defnum{obj/bind/pev/hit}{0.286} \defnum{obj/bind/pev/enr}{1.54}  \defnum{obj/bind/pev/auc}{0.574}
\defnum{obj/stab/pev/hit}{0.546} \defnum{obj/stab/pev/enr}{12.14} \defnum{obj/stab/pev/auc}{0.864}
\defnum{obj/bind/zs/wells}{11.4} \defnum{obj/stab/zs/wells}{80.6}

\defnum{tr/orc/s1}{1.000}  \defnum{tr/orc/auc}{1.000}
\defnum{tr/pev/s1}{0.502}  \defnum{tr/pev/auc}{0.575}
\defnum{tr/pm/s1}{0.475}   \defnum{tr/pm/auc}{0.549}
\defnum{tr/co/s1}{0.437}   \defnum{tr/co/auc}{0.570}
\defnum{tr/zs/s1}{0.425}   \defnum{tr/zs/auc}{0.549}
\defnum{tr/rnd/s1}{0.386}  \defnum{tr/rnd/auc}{0.517}
\defnum{tr/reg/s1}{0.357}  \defnum{tr/reg/auc}{0.563}
\defnum{tr/gain_rnd}{30}   \defnum{tr/gain_reg}{41}   \defnum{tr/gain_zs}{18}   \defnum{tr/gain_pm}{6}
\defnum{tr/d_reg}{0.145}   \defnum{tr/d_co}{0.065}    \defnum{tr/d_pm}{0.027}

\defnum{rel/imp_before}{0.182}    \defnum{rel/imp_after}{0.096}
\defnum{rel/choice_before}{0.203} \defnum{rel/choice_after}{0.114}
\defnum{rel/imp_pct}{47}          \defnum{rel/choice_pct}{44}

\defnum{gb/ceil/16}{2.08} \defnum{gb/ceil/32}{5.57} \defnum{gb/ceil/64}{6.48}
\defnum{gb/pev/16}{1.43}  \defnum{gb/pev/32}{3.61}  \defnum{gb/pev/64}{5.33}
\defnum{gb/ridge/16}{1.03} \defnum{gb/ridge/32}{2.74} \defnum{gb/ridge/64}{4.86}
\defnum{gb/gp/16}{0.76}   \defnum{gb/gp/32}{2.17}   \defnum{gb/gp/64}{5.00}
\defnum{gb/pm/16}{1.24}   \defnum{gb/pm/32}{2.99}   \defnum{gb/pm/64}{4.41}
\defnum{gb/reg/16}{0.83}  \defnum{gb/reg/32}{2.35}  \defnum{gb/reg/64}{4.16}
\defnum{gb/co/16}{1.18}   \defnum{gb/co/32}{2.36}   \defnum{gb/co/64}{3.30}
\defnum{gb/rnd/16}{0.85}  \defnum{gb/rnd/32}{2.17}  \defnum{gb/rnd/64}{3.97}
\defnum{gb/lead/16}{39}   \defnum{gb/lead/32}{32}   \defnum{gb/lead/64}{10}
\defnum{gb/share/pev/16}{47} \defnum{gb/share/pev/32}{42} \defnum{gb/share/pev/64}{54}
\defnum{gb/share/ridge/16}{15} \defnum{gb/share/ridge/32}{17} \defnum{gb/share/ridge/64}{35}
\defnum{gb/share/gp/64}{41}
\defnum{gb/d_reg64}{1.17}
\defnum{gb/n_init}{50}  \defnum{gb/budget}{64}  \defnum{gb/batch}{8}  \defnum{gb/legal}{76}

\defnum{fr/pev/p}{0.814}  \defnum{fr/pev/s}{17{,}344}
\defnum{fr/co/p}{0.799}   \defnum{fr/co/s}{17{,}244}
\defnum{fr/reg/p}{0.773}  \defnum{fr/reg/s}{17{,}216}
\defnum{fr/pm/p}{0.766}   \defnum{fr/pm/s}{21}
\defnum{fr/rnd/p}{0.661}
\defnum{fr/speedup}{800}
\defnum{fr/pev/req}{0.06}  \defnum{fr/pm/req}{48}
\defnum{fr/d_pm}{0.048}   \defnum{fr/d_reg}{0.041}   \defnum{fr/d_co}{0.015}

\defnum{data/eval_variants}{214{,}378}  \defnum{data/eval_fam}{165}  \defnum{data/eval_ctx}{157}
\defnum{data/public}{nine}  \defnum{data/train_public}{eight}
\defnum{eval/wells}{10}     \defnum{eval/min_cand}{100}

\definecolor{bg}{HTML}{E7E7FE}
\definecolor{accent-bright}{HTML}{6B67EE}

\title{\fontsize{0.55cm}{0.66cm}\selectfont A System One Approach to General Protein Evolution}

\author[1]{Yibo Wen}
\author[1]{Han Liu}

\affiliation[1]{Northwestern University}
\abstract{Protein engineering requires deciding which mutation to test next and when to stop.
Existing approaches often require target-family measurements, encode candidates separately, or return scores that are not calibrated to the assay.
We introduce \textbf{\Pev}, a general mutation policy that explores a System One approach to protein evolution: learning from experiments across families to make decisions in a single parallel query.
Given a parent sequence and an assay context, \Pev encodes the parent once and returns calibrated probabilities for choosing an edit or stopping, together with each edit's probability of improving the assay outcome.
One policy covers peptides, small folded domains, and antibodies across binding and stability objectives, without fitting model weights to the target family.
In ten-well screens on held-out families, \Pev finds \num{disc/hit_x_rnd}$\times$ as many beneficial mutations as random selection and \num{disc/hit_x_zs}$\times$ as many as zero-shot ESM-2.
It also outperforms a ridge surrogate refitted on target measurements in closed-loop search on the held-out GB1 landscape.
Post-hoc calibration improves probability reliability without changing decisions, while shared encoding provides over \num{fr/speedup}$\times$ the throughput of per-mutant encoding at 1{,}000 candidates per request.
These results suggest that a System One approach could support efficient experimental decisions across protein families, with \Pev providing an initial demonstration.}
\correspondence{\email{hanliu@northwestern.edu}}
\begin{document}
\maketitle

\renewcommand{\footnoterule}{%
  \kern -3pt
  \hrule width \linewidth
  \kern 2.6pt
}
\vspace{-0.2cm}


\begin{figure}[h!]
  \vspace{0.4cm}
  \centering
  \includegraphics[width=\linewidth]{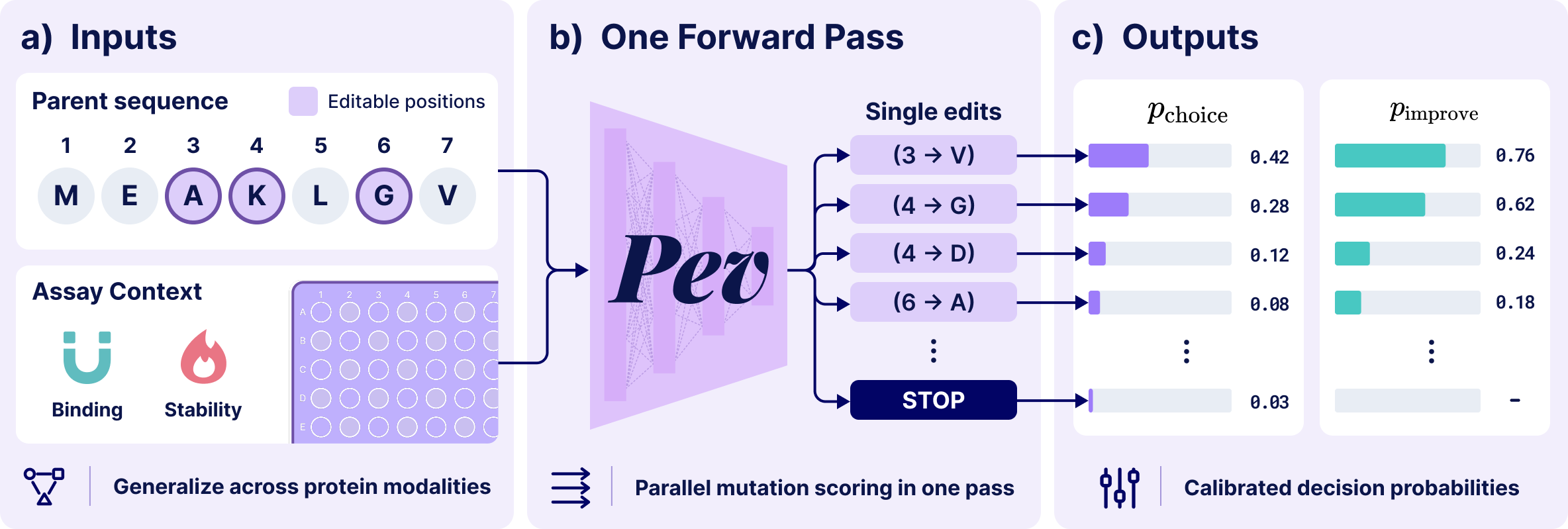}
  \caption{\textbf{Overview of \Pev.} Given a parent sequence and an assay context, \Pev reuses one parent encoding to score all offered substitutions and \STOP in parallel. Calibrated choice and improvement probabilities guide which edit to test and when to stop, supporting a System One approach to general protein evolution.}
  \label{fig:overview}
  \vspace{-0.3cm}
\end{figure}

\section{Introduction}\label{sec:intro}

Protein engineering requires a sequence of experimental decisions under a limited assay budget.
Directed evolution proceeds through rounds of mutation and selection \citep{romero2009exploring,packer2015methods}, with each round requiring a choice of which variants to test and whether further experiments are worthwhile.
These decisions are difficult because the effect of a mutation depends on the surrounding sequence \citep{starr2016epistasis,wu2016adaptation}.
Deep mutational scanning measures thousands of substitutions at once \citep{fowler2014dms}, but each scan serves one target and objective, so a new campaign still spends its budget a few variants at a time.
Effective guidance can nevertheless make small experimental budgets productive: language-model guidance enabled affinity maturation of seven antibodies using at most twenty variants per antibody over two rounds \citep{hie2024antibodies}.
A policy that transfers knowledge across protein families and assay objectives could help new campaigns make informed choices from their first round.

Existing approaches provide useful predictions but leave a gap between scoring variants and making assay-specific decisions.
Family-specific surrogates learn from measurements of the target family \citep{yang2019mlde,wu2019machine,biswas2021low,jiang2025evolvepro,yang2025alde}, using methods such as Gaussian processes and adaptive optimization to guide subsequent experiments \citep{romero2013gp,hie2022adaptive}.
Evolutionary information and pretrained representations improve learning when measurements are scarce \citep{wittmann2021informed,hsu2022learning}, but these surrogates still require target-family data.
Zero-shot protein language models can score variants without such measurements \citep{meier2021language,notin2022tranception,notin2023proteingym}, although their scores are not directly calibrated to assay outcomes and their accuracy can depend on their preference for the wild-type sequence \citep{gordon2025preference}.
Preference alignment adapts generative models to experimental or energy-based objectives \citep{rafailov2023dpo,widatalla2026proteindpo,wen2025alignab}, but does not by itself provide calibrated probabilities for choosing among offered edits or stopping.

Together, these limitations motivate a decision model that transfers across protein families and expresses its predictions as assay-relevant probabilities.
We approach protein evolution from the System One perspective: learning to return a structured decision in a single parallel query.
Jev \citep{typesafe2026jev} exemplifies this approach by mapping structured state to typed outputs with calibrated probabilities.
Protein engineering offers a natural decision structure: a parent sequence, an assay context, and a finite set of candidate edits.
A zero-shot language model also reads the parent once, so a single pass is not what sets a decision model apart; what sets it apart is learning experimental decisions from measured outcomes across families, weighing every offered edit against \STOP in one query, and reporting probabilities calibrated to assay events.
Scoring every candidate from the shared parent representation also spares the backbone pass that encoding each mutant separately needs for every candidate.

We introduce \textbf{\Pev}, a general mutation policy that selects an edit or \STOP in one such query.
Given a parent sequence and an assay context, it encodes the parent once and returns a choice distribution $\pchoice$ over the offered edits and \STOP, and an improvement probability $\pimp$ for each substitution.
\Pev trains approximately 1.6 million parameters with a frozen ESM-2 backbone \citep{lin2023esm2}, learning from measured outcomes across families rather than fitting weights to the family being screened.
Post-hoc calibration uses separate held-out families and preserves action choices and within-panel rankings.

We summarize our \textbf{main contributions} as follows:
\begin{enumerate}[label=\arabic*.,itemsep=2pt]
  \item \textbf{A System One decision model.}
  \Pev formulates protein evolution as choosing among offered edits and \STOP, scoring all actions in one parallel query from a shared parent encoding.

  \item \textbf{Calibrated decision probabilities.}
  Separate choice and improvement outputs distinguish which action is
  best among those offered from whether an edit improves the parent.

  \item \textbf{Transfer across protein families.}
  A single policy learns across peptides, small folded domains, and antibodies under binding and stability objectives, without any family-specific fine-tuning.

\end{enumerate}

We evaluate \Pev on held-out protein families and a held-out fitness landscape, demonstrating improved experimental selection over the evaluated baselines and substantially higher scoring throughput than per-mutant encoding.
Further discussion of related work is provided in App.~\ref{app:related}.
\section{Problem Formulation}
\label{sec:problem}

This section describes the decision that \Pev makes in each round of protein engineering.
\Cref{sec:setting} introduces the parent sequence, the assay context and the edits on offer.
\Cref{sec:utility} defines how an edit is judged against the parent once its outcome is measured.
\Cref{sec:outputs} specifies the two probabilities that \Pev returns for every decision and how they are used to select edits and to stop.

\subsection{Decision Setting}\label{sec:setting}

\paragraph{Parent and context.}
Each decision concerns a parent sequence in an assay context.
The parent $x\in\calA^{L}$ is a protein sequence of length $L$ over the twenty standard amino acids $\calA$, and it belongs to a protein family.
The context $c$ states the objective, binding or stability, and the readout, such as $\Delta\Delta G$, melting temperature $T_m$, affinity, or a binding signal.
For binding it may also include the sequences of a target and a partner.
Within a context, a measured variant $x'$ has an outcome $y_c(x')$, oriented so that higher is better, and a replicate standard deviation $s_c(x')$ when replicates exist.
Outcomes are compared only within the same context.

\paragraph{Offered edits.}
The policy chooses among single substitutions.
An edit $a=(i,r)$ replaces the residue at position $i$ with a different amino acid $r$, and $x\oplus a$ denotes the edited sequence.
The offered panel $\calO(x,c)$ is the set of edits available to the policy.
In a retrospective screen it contains the substitutions that were measured for the parent, so unmeasured mutations are left out rather than counted as misses.
In a campaign it contains the substitutions that have not yet been assayed.
The policy may also choose \STOP, which declines the offered edits and retains the best measured sequence.
Together, the offered edits and \STOP form the action set $\calO(x,c)\cup\{\STOP\}$.

\subsection{Judging an Edit}\label{sec:utility}

\paragraph{Utility.}
An edit is judged by how much it changes the measured outcome.
Its utility is the gain over the parent, and its noise scale combines the replicate standard deviations of the two sequences:
\begin{equation}
  u_a=y_c(x\oplus a)-y_c(x),\qquad
  \sigma_a=\sqrt{s_c(x\oplus a)^2+s_c(x)^2}.
  \label{eq:utility}
\end{equation}
\STOP leaves the parent unchanged, so $u_{\STOP}=0$ and $\sigma_{\STOP}=0$.

\paragraph{Improving and beneficial edits.}
Two events are used throughout the paper.
An edit \emph{improves} the parent when $u_a>0$, which is the event forecast by the improvement probability below.
It is \emph{beneficial} when its gain also exceeds the noise scale and a readout-specific effect floor $\delta_\rho$:
\begin{equation}
  u_a>\max\big(\sigma_a,\ \delta_\rho\big).
  \label{eq:beneficial}
\end{equation}
The floor is fixed for each readout before any model is scored (App.~\ref{app:defs}).
For $\Delta\Delta G$, $T_m$, and affinity it is a conventional effect size, and for binding signals it is half the robust spread of the gains measured in the panel.
Beneficial edits define a hit in the discovery screen (\cref{sec:discovery}).

\paragraph{Best action.}
Measurement noise also means that the best action is not always unique.
We take the best action $a^\star=\argmax_j u_j$ over $\calO(x,c)\cup\{\STOP\}$ and collect in the target set $\mathcal T$ every action whose utility is within noise of the best utility $u_{a^\star}$:
\begin{equation}
  \mathcal T=
  \Big\{\,j:\ u_{a^\star}-u_j
  \le\sqrt{\sigma_j^2+\sigma_{a^\star}^2}\,\Big\}.
  \label{eq:target}
\end{equation}
Since $u_{\STOP}=\sigma_{\STOP}=0$, \STOP belongs to $\mathcal T$ exactly when the highest-gain edit does not exceed the parent beyond its own noise scale, so \STOP is the right choice when no offered edit is reliably better than the parent.

\subsection{Policy Outputs}\label{sec:outputs}

\paragraph{Two probabilities.}
Given the parent, the context and the offered panel, the policy returns two kinds of probability:
\begin{equation}
  \pchoice\in\Delta\big(\calO\cup\{\STOP\}\big),\qquad
  \pimp\in[0,1]^{\calO}.
  \label{eq:policy}
\end{equation}
The choice distribution $\pchoice$ gives each action the probability of being the best on offer, with the probability shared among the actions tied within noise.
It includes \STOP and sums to one.
The improvement probability $\pimp$ gives each edit the probability of improving the parent, $u_a>0$.
Several edits can improve the same parent, so these probabilities need not sum to one.
The two outputs answer different questions: $\pchoice$ sets the edits and \STOP against each other, while $\pimp$ asks of each edit alone whether it is worth trying.

\paragraph{Use in practice.}
The outputs serve both selection and ranking.
The policy selects $\hat a=\argmax\pchoice$, which is an edit or \STOP.
A ten-well screen assays the ten edits with the highest choice probabilities, and a campaign orders each batch by improvement probability.
\STOP carries one meaning in each setting.
In a one-step evaluation it declines to spend the assay, since no offered edit is reliably better than the parent, and in a campaign it retires the current parent once the batch in hand has been assayed, so the search continues from another parent until the budget is spent.
\Cref{sec:method} describes how \Pev computes both outputs from one encoding of the parent.


\section{\texorpdfstring{\Pev}{Pev} for General Protein Evolution}\label{sec:method}

\Pev turns the decision of \cref{sec:problem} into a single forward pass over the parent.
Three design choices define it.
Parent reuse scores every candidate from one encoding of the parent, the decision objective trains the choice distribution on which action is best in place of a regression onto each edit's gain, and the improvement loss gives every edit its own probability of improving the parent.
\Cref{sec:encoding,sec:heads} describe how one encoding yields both outputs, \cref{sec:training} explains how a single policy is trained on measured scans from many families, and \cref{sec:calibfit} adds post-hoc calibration.
\Cref{sec:ablations} removes each design choice in turn.

\subsection{One Encoding for Every Edit}\label{sec:encoding}

\paragraph{Frozen backbone.}
\Pev encodes the parent once and scores every offered edit from that encoding.
The backbone is ESM-2 650M \citep{lin2023esm2}, kept frozen, and its encoding $h=\mathrm{Enc}(x)$ holds one state per residue of the parent.
\Pev trains only the context and residue embeddings and two small networks on top of the backbone, a candidate multilayer perceptron (MLP) and a \STOP head, which together hold about 1.6 million parameters stored in 6\,MB.

\paragraph{Candidate features.}
Each candidate edit $a=(i,r)$ is described by six concatenated parts:
\begin{equation}
  \HL{\ \phi_a=\big[\,h_i;\ \bar h;\ e_r;\ q_c;\ e_{x_i};\ \ell_a\,\big]\ },
\end{equation}
\begin{equation}
  \ell_a=\big[\log p_{\mathrm{LM}}(r\mid x,i)-\log p_{\mathrm{LM}}(x_i\mid x,i),\ \log p_{\mathrm{LM}}(x_i\mid x,i)\big].
\end{equation}
Here $h_i$ is the parent's state at the edited position, $\bar h$ the pooled parent state, and $e_r$ and $e_{x_i}$ embed the new and the original residue.
The context embedding $q_c$ covers the objective, the readout and the projected target and partner states, which come from the same frozen backbone, once per sequence.
The vector $\ell_a$ holds the backbone's own language-model log-odds for the edit, read from the unmasked parent at no extra cost, and its first entry is the score that the zero-shot ESM-2 baseline ranks by.

\paragraph{Parent reuse.}
All candidate features are assembled from a shared parent encoding, the assay context and edit-specific parts that need no encoder of their own, so a decision over $N$ candidates costs one backbone pass plus a light head per candidate (\cref{prop:cost}).
Encoding every mutant separately would cost one pass per candidate.
The per-mutant baseline of \cref{sec:experiments} differs from \Pev in exactly this one place: $h_i$ and $\bar h$ come from encoding $x\oplus a$ separately, and its heads, conditioning and training are the same as in \Pev.

\subsection{Two Outputs for Two Questions}\label{sec:heads}

\paragraph{Two heads.}
The candidate MLP has two hidden layers and maps each candidate to a choice logit and an improvement logit, $(z_a,v_a)=\mathrm{MLP}(\phi_a)$.
A separate head with one hidden layer maps the pooled state and the context to the \STOP logit $z_{\STOP}$.
With the choice logits $z=\big([z_a]_{a\in\calO},\,z_{\STOP}\big)$, the two outputs are
\begin{equation}
  \pchoice=\softmax(z),\qquad \pimp[a]=\sigma(v_a),
\end{equation}
where $\sigma$ is the logistic sigmoid.
The choice head is a softmax over the edits and \STOP, so it sums to one, and the improvement head is an independent sigmoid for each edit, so its outputs do not sum to one.
Neither substitutes for the other: $\pchoice$ sets the edits and \STOP against each other, while $\pimp$ asks of each edit alone whether it improves the parent.
Because the output space is the closed set of offered edits and \STOP (\cref{fig:overview}), nothing is generated, and every output is a valid decision that a laboratory can act on.

\subsection{Training on Wild-type Scans}\label{sec:training}

\paragraph{Episodes.}
\Pev trains on wild-type scans, the kind of panel that a deep mutational scan produces.
An episode is a measured reference plus up to 64 of its measured single substitutions, resampled each time the reference is drawn.
It has the same form as a discovery-screen panel, so one encoder pass covers every candidate in it.
Because the options have the same shape in every family, a set of single substitutions plus \STOP, scans from many families train one policy.

\paragraph{Targets and losses.}
Each head is trained on a target that respects measurement noise.
The choice target $p^\ast$ is uniform on the target set $\mathcal T$ of \cref{eq:target}, which holds the best action and every action within noise of it.
The improvement target $y^\ast_a$ is a noise-aware soft label for the event $u_a>0$ (App.~\ref{app:defs}).
The decision objective is the cross-entropy of the choice distribution to $p^\ast$, and the improvement loss is the binary cross-entropy (BCE) of each improvement probability to its label, averaged over the offered edits:
\begin{equation}
  \mathcal L_{\mathrm{choice}}=-\sum_j p^\ast_j\log\softmax(z)_j,\qquad
  \mathcal L_{\mathrm{improve}}=\tfrac1{|\calO|}\sum_{a\in\calO}\mathrm{BCE}\big(\sigma(v_a),y^\ast_a\big).
\end{equation}

\paragraph{Objective.}
The two terms are combined as $\mathcal L(\theta)=\E_\omega[w_\omega(\mathcal L_{\mathrm{choice}}+\lambda\,\mathcal L_{\mathrm{improve}})]$ with $\lambda=1$, where $\theta$ holds the trained parameters, $\omega$ is an episode, and the weight $w_\omega\le1$ lowers the contribution of an episode whose best action lies within noise (App.~\ref{app:defs}).
Both terms are strictly proper scoring rules, so each head is minimized only at the weighted conditional mean of its own training target (\cref{prop:proper}), the noise-aware soft label rather than the event itself.
Properness therefore fixes what each head estimates, and the reported numbers are matched to the frequencies of a new assay by the calibration of \cref{sec:calibfit}.
Censored measurements never enter either loss, and losing candidates stay in the denominator of the choice term.
Training details are given in App.~\ref{app:train}.

\subsection{Post-hoc Calibration}\label{sec:calibfit}

\paragraph{Calibration maps.}
Training fits the logits to noise-aware targets, so the raw probabilities need not match the frequencies observed in a new assay.
Post-hoc calibration fits three parameters for each objective on held-out calibration families, against hard labels for the two events rather than the soft targets used in training (App.~\ref{app:defs}): a temperature $T>0$ that scales all choice logits within a decision, \STOP included, and an affine map with positive slope $\alpha$ and intercept $\beta$ on the improvement logits.
The calibrated outputs are
\begin{equation}
  \pchoice=\softmax(z/T),\qquad
  \pimp[a]=\sigma(\alpha v_a+\beta),
\end{equation}
and training itself uses $T=\alpha=1$ and $\beta=0$.

\paragraph{Decisions are preserved.}
Calibration changes how confident the probabilities are and not which action is chosen.
Both maps are monotone, so they cannot reorder the logits, which is what the following proposition states.

\begin{proposition}[Calibration preserves choices and rankings]\label{prop:calib}
Let $T>0$, $\alpha>0$, and $\beta$ be the calibration parameters for a decision's objective.
Then
\[
  \argmax_j\softmax(z/T)_j=\argmax_j z_j,
\]
and $\sigma(\alpha v_a+\beta)$ orders the offered edits exactly as $v_a$ does.
Thus, with the same tie-breaking rule, calibration preserves the selected action $\hat a$, the within-panel rankings under both outputs, and every metric that depends only on these choices and rankings (App.~\ref{app:defs}).
Metrics that use the probability values themselves, including expected calibration error (ECE), may change.
\end{proposition}

\begin{proof}
Dividing by $T>0$ preserves the order of the choice logits, and softmax preserves their pairwise ordering.
The map $v\mapsto\sigma(\alpha v+\beta)$ is strictly increasing because $\alpha>0$ and the logistic sigmoid is strictly increasing.
Both transformations therefore preserve rankings, including ties.
\end{proof}


\FloatBarrier
\section{Experiments}\label{sec:experiments}

We evaluate whether \Pev's decisions are useful, whether its probabilities can be trusted, and whether one encoding of the parent is efficient.
\Cref{sec:setup} describes the evaluation settings, data and baselines.
\Cref{sec:discovery,sec:transfer} test one-round decisions on held-out protein families, and \cref{sec:closedloop} tests sequential selection on a held-out fitness landscape.
\Cref{sec:calibration} checks that the predicted probabilities are reliable, \cref{sec:frontier} compares selection quality with throughput, and \cref{sec:ablations} examines the contribution of each design choice.

\begin{table}[H]
  \centering
  \caption{\textbf{Evaluation settings.} One-round screens assess candidate ranking and edit-or-stop decisions, and closed-loop campaigns assess sequential batch selection under an assay budget.}
  \label{tab:settings}
  \tablestyle\footnotesize
  \begin{tabularx}{\linewidth}{H Y Y}
    \toprule
    \headrow \hd{} & \hd{One-round screen\quad(\cref{sec:discovery,sec:transfer})} & \hd{Closed-loop campaign\quad(\cref{sec:closedloop})}\\
    \midrule
    Candidates & Measured single substitutions of a reference parent, from full wild-type scans or sparse neighborhoods & All unassayed single edits of the parent, at most \num{gb/legal}\\
    Budget & Ten wells (discovery) or one action (edit-or-stop) & \num{gb/budget} assays, ordered in batches of \num{gb/batch}\\
    Decision & Which edits to prioritize, or whether to select \STOP & Which batch to assay next and which parent to use\\
    Metrics & Hit rate, enrichment, AUROC, wells to hit, Success@1 & Best observed fitness $F_B$, normalized to wild type\\
    Held out & Protein family & Fitness landscape\\
    Population & \num{disc/panels} panels from \num{disc/fam} of the \num{data/eval_fam} held-out families & One landscape, over \num{gb/n_init} paired initializations\\
    Eligibility & At least \num{eval/min_cand} measured substitutions of the reference, and one beneficial edit & Single edits of the parent, not yet assayed\\
    \bottomrule
  \end{tabularx}
\end{table}

\subsection{Experimental Setup}\label{sec:setup}

\paragraph{Settings.}
The experiments use two evaluation settings.
In a one-round screen, \Pev ranks the measured single substitutions of a parent and is scored against their measured outcomes, either on a ten-well budget (\cref{sec:discovery}) or one decision at a time (\cref{sec:transfer}).
In a closed-loop campaign (\cref{sec:closedloop}), \Pev orders batches of unassayed edits under an assay budget, and the campaign is scored by the best sequence it reaches.
In neither setting are \Pev's weights fitted to the target: a one-round screen gives it no measurement of the family at all, and in a campaign the measurements in hand only decide which parent it works from.
The two settings also answer different questions, one about the quality of a single ranking and one about where a sequence of rankings arrives.

\paragraph{Data and splits.}
All evaluations use protein families or a fitness landscape that were held out before training (\cref{tab:settings}).
The corpus draws on \num{data/public} public sources: \num{data/train_public} supply the training and evaluation data, and GB1 is reserved for the closed-loop experiment (App.~\ref{app:data}).
Splits are made by family, and a family held out for one property remains held out for every property.
The evaluation split contains \num{data/eval_fam} families and \num{data/eval_variants} variants, and both one-round screens draw on it: the discovery benchmark takes \num{disc/panels} panels from the \num{disc/fam} of those families that its rule admits, and the transfer test takes the sparse neighborhoods of the same split.

\paragraph{Baselines.}
We compare \Pev with methods that need no target-family measurements, a surrogate that learns from campaign measurements, and variants that isolate its design choices.
Random selection and zero-shot ESM-2 are the baselines in every screen, where zero-shot ESM-2 ranks edits by the log-odds of the backbone and never selects \STOP.
The discovery screen includes ESM-1v \citep{meier2021language}, a language model trained for variant effect prediction, scored by masked marginals over its five published checkpoints.
The closed-loop experiment adds two comparators refitted on the target's observed assays: a ridge surrogate on one-hot site features, and Bayesian optimization with a Gaussian process over the same features, using a squared-exponential kernel and ranked by expected improvement.
Three variants each remove one design choice of \cref{sec:method}: the per-mutant variant encodes every mutant separately instead of sharing the parent encoding, the regression variant replaces the decision objective with a point prediction trained by mean squared error, and the choice-only variant sets $\lambda=0$ to remove the improvement loss.

\paragraph{Metrics and statistics.}
Each subsection defines its metrics where they are first used, and full definitions are given in App.~\ref{app:defs}.
Screen comparisons are paired by family, with uncertainty estimated by a family-clustered bootstrap, and closed-loop comparisons are paired by initialization.

\subsection{Finding Beneficial Mutations in Held-out Families}\label{sec:discovery}

\paragraph{Setup.}
The discovery screen asks how well \Pev finds beneficial edits when it can test only ten.
A panel is one measured reference protein with its measured single substitutions in one assay context, and \Pev orders each panel with its choice head.
An edit counts as a hit when it is beneficial in the sense of \cref{eq:beneficial}.
Hit rate@10 is the share of the ten selected edits that are hits, and enrichment@10 divides it by the panel's share of beneficial edits, so that random selection scores one in expectation.
We also report the AUROC of the full ranking and the rank of the first beneficial edit, which we call wells to first hit.
The panel list is frozen before anything is scored on it.

\begin{table}[H]
  \centering
  \caption{\textbf{Discovery on held-out families.} Means over \num{disc/panels} panels from \num{disc/fam} families (\num{cls/dom/fam} domain, \num{cls/pep/fam} peptide and \num{cls/ab/fam} antibody) with \num{disc/cand} measured candidates and ten wells per panel. Hit@10 and Enr.@10 are hit rate and enrichment at ten wells, and wells to hit is the rank of the first beneficial edit. Perfect ranking is the ceiling that each panel's prevalence sets, and the best of the four methods in each block is in bold.}
  \label{tab:discovery}
  \tablestyle\footnotesize\setlength{\tabcolsep}{3.0pt}
  \begin{tabularx}{\linewidth}{Y cccc ccc ccc ccc}
    \toprule
    \headrow \hd{} & \hdgroup{4}{All, \num{disc/panels} panels} & \hdgroup{3}{Domain, \num{cls/dom/panels} panels}
      & \hdgroup{3}{Peptide, \num{cls/pep/panels} panels} & \hdgroup{3}{Antibody, \num{cls/ab/panels} panels}\\
    \cmidrule(lr){2-5}\cmidrule(lr){6-8}\cmidrule(lr){9-11}\cmidrule(lr){12-14}
    \headrow \hd{Method}
      & \hdtwo{Hit\\@10\up} & \hdtwo{Enr.\\@10\up} & \hdtwo{AUROC\\\up} & \hdtwo{Wells\\to hit\dn}
      & \hdtwo{Hit\\@10\up} & \hdtwo{Enr.\\@10\up} & \hdtwo{AUROC\\\up}
      & \hdtwo{Hit\\@10\up} & \hdtwo{Enr.\\@10\up} & \hdtwo{AUROC\\\up}
      & \hdtwo{Hit\\@10\up} & \hdtwo{Enr.\\@10\up} & \hdtwo{AUROC\\\up}\\
    \midrule
    \ceilrow{Perfect ranking} & \ceilrow{\num{disc/orc/hit}} & \ceilrow{\num{disc/orc/enr}$\times$} & \ceilrow{\num{disc/orc/auc}} & \ceilrow{\num{disc/orc/wells}}
      & \ceilrow{\num{cls/dom/orc/hit}} & \ceilrow{\num{cls/dom/orc/enr}$\times$} & \ceilrow{\num{cls/dom/orc/auc}}
      & \ceilrow{\num{cls/pep/orc/hit}} & \ceilrow{\num{cls/pep/orc/enr}$\times$} & \ceilrow{\num{cls/pep/orc/auc}}
      & \ceilrow{\num{cls/ab/orc/hit}} & \ceilrow{\num{cls/ab/orc/enr}$\times$} & \ceilrow{\num{cls/ab/orc/auc}}\\
    \ours \textbf{\Pev} & \best{\num{disc/pev/hit}} & \best{\num{disc/pev/enr}$\times$} & \best{\num{disc/pev/auc}} & \best{\num{disc/pev/wells}}
      & \best{\num{cls/dom/pev/hit}} & \best{\num{cls/dom/pev/enr}$\times$} & \best{\num{cls/dom/pev/auc}}
      & \best{\num{cls/pep/pev/hit}} & \best{\num{cls/pep/pev/enr}$\times$} & \best{\num{cls/pep/pev/auc}}
      & \best{\num{cls/ab/pev/hit}} & \num{cls/ab/pev/enr}$\times$ & \num{cls/ab/pev/auc}\\
    ESM-2 & \num{disc/zs/hit} & \num{disc/zs/enr}$\times$ & \num{disc/zs/auc} & \num{disc/zs/wells}
      & \num{cls/dom/zs/hit} & \num{cls/dom/zs/enr}$\times$ & \num{cls/dom/zs/auc}
      & \num{cls/pep/zs/hit} & \num{cls/pep/zs/enr}$\times$ & \num{cls/pep/zs/auc}
      & \num{cls/ab/zs/hit} & \best{\num{cls/ab/zs/enr}$\times$} & \best{\num{cls/ab/zs/auc}}\\
    ESM-1v & \num{disc/e1v/hit} & \num{disc/e1v/enr}$\times$ & \num{disc/e1v/auc} & \num{disc/e1v/wells}
      & \num{cls/dom/e1v/hit} & \num{cls/dom/e1v/enr}$\times$ & \num{cls/dom/e1v/auc}
      & \num{cls/pep/e1v/hit} & \num{cls/pep/e1v/enr}$\times$ & \num{cls/pep/e1v/auc}
      & \num{cls/ab/e1v/hit} & \num{cls/ab/e1v/enr}$\times$ & \num{cls/ab/e1v/auc}\\
    Random & \num{disc/rnd/hit} & \num{disc/rnd/enr}$\times$ & \num{disc/rnd/auc} & \num{disc/rnd/wells}
      & \num{cls/dom/rnd/hit} & \num{cls/dom/rnd/enr}$\times$ & \num{cls/dom/rnd/auc}
      & \num{cls/pep/rnd/hit} & \num{cls/pep/rnd/enr}$\times$ & \num{cls/pep/rnd/auc}
      & \num{cls/ab/rnd/hit} & \num{cls/ab/rnd/enr}$\times$ & \num{cls/ab/rnd/auc}\\
    \bottomrule
  \end{tabularx}
\end{table}


\paragraph{Overall results.}
\Pev finds beneficial edits far more often than every baseline (\cref{tab:discovery}).
Its hit rate at ten wells is \num{disc/pev/hit}, against \num{disc/zs/hit} for zero-shot ESM-2, \num{disc/e1v/hit} for ESM-1v and \num{disc/rnd/hit} for random selection, so the same ten wells hold \num{disc/hit_x_zs}$\times$, \num{disc/hit_x_e1v}$\times$ and \num{disc/hit_x_rnd}$\times$ as many beneficial edits.
ESM-1v is built for variant effects and still trails zero-shot ESM-2 here: both score how likely a substitution is, not whether it improves the parent.
One family can supply several panels, so averaging within each family first, the paired margins in hit rate are \num{disc/m_zs/hit} over zero-shot ESM-2 ([\num{disc/m_zs/hit_lo}, \num{disc/m_zs/hit_hi}], 95\%) and \num{disc/m_e1v/hit} over ESM-1v ([\num{disc/m_e1v/hit_lo}, \num{disc/m_e1v/hit_hi}]).

\wrapguard{18}
\begin{wrapfigure}[16]{r}{0.40\textwidth}
  \vspace{-11pt}
  \centering
  \captionsetup{justification=raggedright}\hyphenpenalty=10000
  \includegraphics[width=\linewidth]{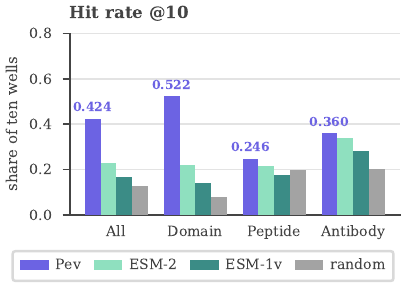}
  \caption{\textbf{\Pev leads in hit rate overall and across sequence classes.} Hit rate@10 is the fraction of ten selected edits that are beneficial.}
  \label{fig:discovery}
\end{wrapfigure}

\paragraph{Sequence classes.}
\Pev's discovery gains vary by sequence class (\cref{fig:discovery}).
They are largest on domains, with a hit rate of \num{cls/dom/pev/hit} across \num{cls/dom/fam} held-out families, versus \num{cls/dom/zs/hit} for zero-shot ESM-2 and \num{cls/dom/rnd/hit} for random selection.
The paired margin over random is \num{cls/dom/m_rnd/hit} (95\% bootstrap interval [\num{cls/dom/m_rnd/hit_lo}, \num{cls/dom/m_rnd/hit_hi}]).
On peptides, \Pev leads zero-shot ESM-2 in hit rate, enrichment, and AUROC (\num{cls/pep/pev/hit} versus \num{cls/pep/zs/hit}, \num{cls/pep/pev/enr}$\times$ versus \num{cls/pep/zs/enr}$\times$, and \num{cls/pep/pev/auc} versus \num{cls/pep/zs/auc}).
On the \num{cls/ab/panels} antibody panels, \Pev has higher hit rate (\num{cls/ab/pev/hit} versus \num{cls/ab/zs/hit}), while zero-shot ESM-2 has higher enrichment (\num{cls/ab/zs/enr}$\times$ versus \num{cls/ab/pev/enr}$\times$) and AUROC (\num{cls/ab/zs/auc} versus \num{cls/ab/pev/auc}).

\paragraph{Ranking and first hit.}
\Pev achieves enrichment \num{disc/pev/enr}$\times$ and AUROC \num{disc/pev/auc} (\cref{tab:discovery}), versus \num{disc/zs/enr}$\times$ and \num{disc/zs/auc} for zero-shot ESM-2.
Its first beneficial edit ranks \num{disc/pev/wells} on average, versus \num{disc/zs/wells} for ESM-2, \num{disc/e1v/wells} for ESM-1v and \num{disc/rnd/wells} for random selection.
Full-panel ranks can exceed the ten-well budget of the screen.

\wrapguard{16}
\subsection{Transfer to Sparse Measured Neighborhoods}\label{sec:transfer}

\begin{wraptable}[13]{r}{0.36\textwidth}
  \vspace{-10pt}
  \centering
  \captionsetup{justification=raggedright}\hyphenpenalty=10000
  \caption{\textbf{Transfer to sparse neighborhoods.} \Pev scores highest on both Success@1 and AUROC.}
  \label{tab:transfer}
  \tablestyle\setlength{\tabcolsep}{3.5pt}\renewcommand{\arraystretch}{1.25}
  \begin{tabularx}{\linewidth}{Y rr}
    \toprule
    \headrow \hd{Method} & \hdr{Success@1\up} & \hdr{AUROC\up}\\
    \midrule
    \ceilrow{Perfect decisions} & \ceilrow{\num{tr/orc/s1}} & \ceilrow{\num{tr/orc/auc}}\\
    \ours \textbf{\Pev} & \best{\num{tr/pev/s1}} & \best{\num{tr/pev/auc}}\\
    Zero-shot ESM-2 & \num{tr/zs/s1} & \num{tr/zs/auc}\\
    Per-mutant & \num{tr/pm/s1} & \num{tr/pm/auc}\\
    Regression & \num{tr/reg/s1} & \num{tr/reg/auc}\\
    Choice-only & \num{tr/co/s1} & \num{tr/co/auc}\\
    Random & \num{tr/rnd/s1} & \num{tr/rnd/auc}\\
    \bottomrule
  \end{tabularx}
  \vspace{-4pt}
\end{wraptable}

\paragraph{Setup.}
We evaluate \Pev on sparse neighborhoods with a median of sixteen measured single-edit neighbors per parent, roughly half with positive gains.
Most panels contain fewer hits than the ten-well budget, so screening metrics saturate; we instead report two budget-independent metrics.
An edit clears its noise when its gain exceeds the replicate noise, $u_a>\sigma_a$.
Success@1 credits a panel when the selected edit clears its noise or the model chooses \STOP when no offered edit does.
AUROC evaluates the ranking of all offered edits using the same label.
Panels are weighted by class, objective, and family (App.~\ref{app:defs}).

\paragraph{Results.}
\Pev achieves the highest Success@1 and AUROC among the six evaluated methods (\cref{tab:transfer}).
Its Success@1 is \num{tr/pev/s1}, exceeding zero-shot ESM-2 by \num{tr/gain_zs}\%.
Both methods require no measurements from the target family.
However, Success@1 credits correct stopping, which \Pev and its variants support but zero-shot ESM-2 and random selection do not.
AUROC evaluates edit rankings independently of \STOP, with \Pev scoring \num{tr/pev/auc} versus \num{tr/zs/auc} for zero-shot ESM-2.

\begin{figure}[H]
  \centering
  \includegraphics[width=\linewidth]{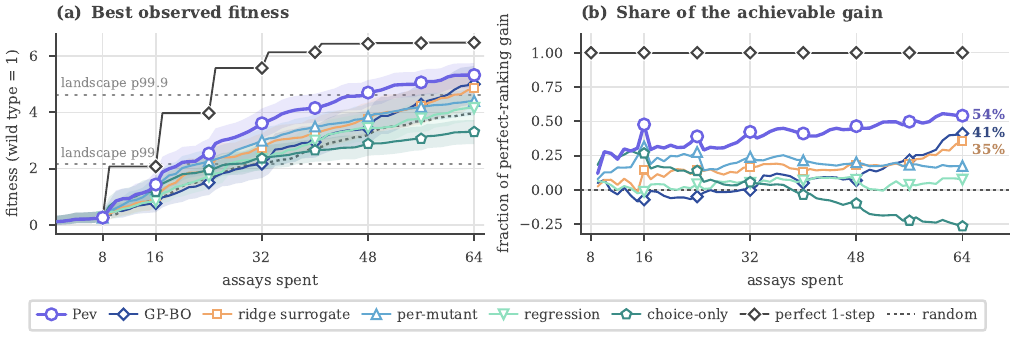}
  \caption{\textbf{Closed-loop campaign on the GB1 four-site landscape.} (a) Mean best observed fitness over \num{gb/n_init} paired initializations, with bootstrap bands, markers every \num{gb/batch} assays, and grey lines at the landscape's 99th and 99.9th percentiles. (b) Each method's fraction of the gap from random selection to perfect one-step ranking across assay budgets.}
  \label{fig:gb1}
\end{figure}

\wrapguard{18}
\subsection{Closed-loop Protein Optimization}\label{sec:closedloop}

\begin{wraptable}[14]{r}{0.37\textwidth}
  \vspace{-8pt}
  \centering
  \captionsetup{justification=raggedright}\hyphenpenalty=10000
  \caption{\textbf{Closed-loop GB1 campaign.} Mean best observed fitness after $B$ assays.}
  \label{tab:gb1}
  \tablestyle\setlength{\tabcolsep}{4pt}\renewcommand{\arraystretch}{1.30}
  \begin{tabularx}{\linewidth}{Y rrr}
    \toprule
    \headrow \hd{Method} & \hdr{$B{=}16$\up} & \hdr{$B{=}32$\up} & \hdr{$B{=}64$\up}\\
    \midrule
    \ceilrow{Perfect ranking} & \ceilrow{\num{gb/ceil/16}} & \ceilrow{\num{gb/ceil/32}} & \ceilrow{\num{gb/ceil/64}}\\
    \ours \textbf{\Pev} & \best{\num{gb/pev/16}} & \best{\num{gb/pev/32}} & \best{\num{gb/pev/64}}\\
    GP-BO & \num{gb/gp/16} & \num{gb/gp/32} & \num{gb/gp/64}\\
    Ridge surrogate & \num{gb/ridge/16} & \num{gb/ridge/32} & \num{gb/ridge/64}\\
    Per-mutant & \num{gb/pm/16} & \num{gb/pm/32} & \num{gb/pm/64}\\
    Regression & \num{gb/reg/16} & \num{gb/reg/32} & \num{gb/reg/64}\\
    Choice-only & \num{gb/co/16} & \num{gb/co/32} & \num{gb/co/64}\\
    Random & \num{gb/rnd/16} & \num{gb/rnd/32} & \num{gb/rnd/64}\\
    \bottomrule
  \end{tabularx}
  \vspace{0pt}
\end{wraptable}

\paragraph{Setup.}
The closed-loop experiment tests whether \Pev's decisions yield better sequences under limited assay budgets. We run campaigns on the held-out GB1 four-site landscape. Each round, the policy selects a batch of \num{gb/batch} from at most \num{gb/legal} unassayed single edits of the current parent. All arms use the full \num{gb/budget}-assay budget, including initialization. \Pev ranks edits by calibrated improvement probability; \STOP retires the parent after its batch, and the campaign continues from another.

\paragraph{Results.}
The ridge surrogate and the Gaussian process are both fitted to the target's own assays and refit after each round, and all arms share initial sets, batch rules, and budgets. Across \num{gb/n_init} paired initializations, \Pev achieves higher best observed fitness $F_B=\max_{t\le B}y(x_t)$ than both at all three budgets ($B=16,32,64$), with wild-type fitness set to one (\cref{tab:gb1}). Its advantage emerges in the earliest batches, when a model fitted to the target has the least training data. At 64 assays, \Pev reaches \num{gb/pev/64} against \num{gb/ridge/64} for the surrogate and \num{gb/gp/64} for the Gaussian process.

\paragraph{Share of the achievable gap.}
\Cref{fig:gb1}b rescales the campaigns of \cref{fig:gb1}a by the gap between random selection and perfect one-step ranking, which assays the \num{gb/batch} best unassayed neighbors in every round.
\Pev closes \num{gb/share/pev/64}\% of that gap by 64 assays, the Gaussian process \num{gb/share/gp/64}\% and the ridge surrogate \num{gb/share/ridge/64}\%.

\subsection{Calibrated Probabilities}\label{sec:calibration}

\paragraph{Metric.}
A decision policy is most useful when its probabilities can be read as probabilities.
We measure this with the expected calibration error (ECE), the weighted mean absolute gap between predicted probability and observed frequency across ten equal-width bins, on test families that are disjoint from the families used to fit the calibration (App.~\ref{app:defs}).
The improvement probability is scored against whether each edit improves the parent, and the choice probability of the top-ranked edit against whether that edit is the best improving edit of its panel.

\vspace{-0.3em}
\begin{figure}[H]
  \centering
  \captionsetup{justification=raggedright}\hyphenpenalty=10000
  \begin{minipage}[t]{0.575\linewidth}
    \centering
    \includegraphics[width=\linewidth]{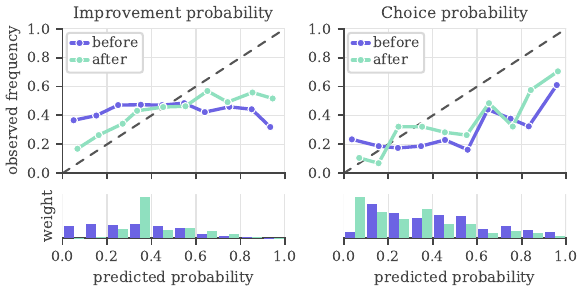}
    \caption{\textbf{Calibration outcomes on test families.} Observed versus predicted probabilities across ten bins, before and after calibration. The dashed diagonal indicates perfect calibration, and bars below show the distribution of predictions across bins.}
    \label{fig:reliability}
  \end{minipage}\hfill
  \begin{minipage}[t]{0.40\linewidth}
    \centering
    \includegraphics[width=\linewidth]{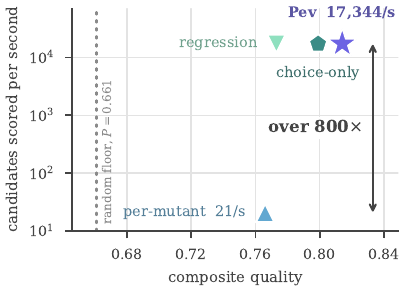}
    \caption{\textbf{Quality against throughput.} \Pev has the best composite quality $P$, and scores over \num{fr/speedup}$\times$ faster than the per-mutant encoder at $N{=}1000$ candidates in one request.}
    \label{fig:frontier}
  \end{minipage}
\end{figure}
\vspace{-0.8em}

\paragraph{Results.}
A score can rank edits well and still overstate how likely they are to help, and post-hoc calibration corrects the second without touching the first.
It roughly halves the calibration error of both probabilities (\cref{fig:reliability}).
The ECE of the improvement probability falls from \num{rel/imp_before} to \num{rel/imp_after}, a fall of \num{rel/imp_pct}\%.
For the choice probability of the top-ranked edit it falls from \num{rel/choice_before} to \num{rel/choice_after}, a fall of \num{rel/choice_pct}\%.
Calibration changes no decision (\cref{prop:calib}), so only the probabilities that \Pev reports are affected.

\subsection{The Quality and Throughput Frontier}\label{sec:frontier}

\paragraph{Setup.}
We test whether sharing a parent encoding improves scoring efficiency without sacrificing selection quality.
We compare \Pev with three variants trained under the same protocol.
The matched per-mutant variant differs only in whether candidates reuse the parent encoding, isolating the effect of this design choice.
Selection quality is measured by $P$, the fraction of a panel's best gain captured by the top eight edits, averaged within each sequence class and then equally across classes (App.~\ref{app:defs}).
Throughput is measured in candidates per second for requests of $N{=}1000$ candidates, with CUDA synchronization (App.~\ref{app:train}).

\paragraph{Results.}
Shared encoding improves throughput while retaining strong selection quality (\cref{fig:frontier}).
\Pev achieves the highest $P$ among the four variants and over \num{fr/speedup}$\times$ the throughput of the matched per-mutant encoder.
It scores \num{fr/pev/s} candidates per second versus \num{fr/pm/s}, while achieving $P=\num{fr/pev/p}$ on held-out parents versus \num{fr/pm/p}.
These results show a Pareto improvement over per-mutant encoding in the evaluated setting: \Pev selects better edits at substantially lower computational cost.
The shared backbone pass also avoids repeating the dominant computation as the candidate set grows (\cref{prop:cost}).

\wrapguard{11}
\subsection{Ablation Studies}\label{sec:ablations}

\begin{wraptable}[9]{r}{0.54\textwidth}
  \vspace{-12pt}
  \centering
  \captionsetup{justification=raggedright}\hyphenpenalty=10000
  \caption{\textbf{Ablations.} Each variant removes one design choice; speed is candidates scored per second.}
  \label{tab:variants_ablation}
  \tablestyle\setlength{\tabcolsep}{4pt}\renewcommand{\arraystretch}{1.12}
  \begin{tabularx}{\linewidth}{Y rrrr}
    \toprule
    \headrow \hd{Method} & \hdr{$P$\up} & \hdr{Speed\up} & \hdr{Success@1\up} & \hdr{Best fitness\up}\\
    \midrule
    \ours \textbf{\Pev} & \best{\num{fr/pev/p}} & \best{\num{fr/pev/s}} & \best{\num{tr/pev/s1}} & \best{\num{gb/pev/64}}\\
    Per-mutant & \num{fr/pm/p} & \num{fr/pm/s} & \num{tr/pm/s1} & \num{gb/pm/64}\\
    Regression & \num{fr/reg/p} & \num{fr/reg/s} & \num{tr/reg/s1} & \num{gb/reg/64}\\
    Choice-only & \num{fr/co/p} & \num{fr/co/s} & \num{tr/co/s1} & \num{gb/co/64}\\
    Random & \num{fr/rnd/p} & n/a & \num{tr/rnd/s1} & \num{gb/rnd/64}\\
    \bottomrule
  \end{tabularx}
  \vspace{-4pt}
\end{wraptable}

\paragraph{Setup.}
We evaluate three matched variants to assess the contributions of parent reuse, the decision objective, and the improvement loss (\cref{tab:variants_ablation}).
The per-mutant variant encodes each candidate separately, the regression variant replaces the decision objective with mean squared error, and the choice-only variant omits the improvement loss.
All other settings are held fixed, with random selection providing a model-free baseline.

\paragraph{Results.}
Each variant scores below \Pev on $P$, Success@1, and best observed fitness at 64 assays.
Shared parent encoding improves $P$ by \num{fr/d_pm} and Success@1 by \num{tr/d_pm} over per-mutant encoding, alongside over \num{fr/speedup}$\times$ higher throughput at $N{=}1000$ candidates.
The decision objective improves these quality metrics over regression by \num{fr/d_reg} and \num{tr/d_reg}, respectively, and best observed fitness at 64 assays by \num{gb/d_reg64}.
Adding the improvement loss raises $P$ by \num{fr/d_co} and Success@1 by \num{tr/d_co} over the choice-only variant.
Together, these comparisons support the contribution of each component in the evaluated settings.

\section{Limitations}\label{sec:limitations}

Our evaluation focuses on choosing which single substitution of a measured parent to test next, or whether to stop.
This shared decision format supports evaluation across protein families and objectives.
Multi-site edits and explicit conditioning on experimental history remain outside the scope of the current policy.

All experiments use previously measured outcomes.
One-round screens assess selection among measured substitutions, while closed-loop campaigns assess sequential decisions within the held-out GB1 landscape.
These settings do not establish performance on unmeasured candidates or across new landscapes.
Coverage also varies by sequence class, with antibodies represented by \num{cls/ab/panels} panels.
Prospective assays, additional landscapes, and broader family coverage would help establish where a shared policy transfers reliably and where further adaptation may be needed.

\section{Conclusion}\label{sec:discussion}

\Pev explores a System One approach to protein evolution: mapping a parent sequence, an assay context, and offered edits to an experimental decision in one parallel query.
A shared parent encoding enables efficient scoring, while calibrated choice and improvement probabilities guide edit selection and stopping.
Results on held-out families and the GB1 landscape suggest that experimental knowledge pooled across families can support useful decisions without fitting model weights to the target family, giving new campaigns an informed starting point.

This approach could extend from single substitutions to multi-site edits, from individual decisions to policies informed by experimental history, and from binding and stability to broader protein functions.
A shared decision interface offers a way to learn jointly from these settings as more measurements become available.
Combining these fast decisions
with feedback from real experiments offers a path toward scalable, closed-loop protein engineering.

\paragraph{Code Availability.}
We release the \Pev model code, weights and frozen benchmark.
The code of the main experiments is at \url{https://github.com/Yibo-Wen/Pev}, the weights are at \url{https://huggingface.co/Yibooooo/Pev}, and the project page is at \url{https://yibow.me/pev}.

\makeatletter
\edef\pev@hyph{\the\hyphenpenalty}
\let\pev@bibitem\bibitem
\renewcommand\bibitem[2][]{\pev@bibitem[{#1}]{#2}%
  \parfillskip=0pt plus 0.6\linewidth\relax \hyphenpenalty=\pev@hyph\relax \tolerance=200\relax
  \@ifundefined{pev@fill@#2}{}{\parfillskip=0pt plus 0.15\linewidth\relax \hyphenpenalty=10000\relax \tolerance=4000\relax}}
\@namedef{pev@fill@notin2023proteinnpt}{}
\@namedef{pev@fill@wu2019machine}{}
\@namedef{pev@fill@guo2017calibration}{}
\@namedef{pev@fill@hie2024antibodies}{}
\makeatother
\addtolength{\bibsep}{0.3pt}
\bibliographystyle{acl_natbib.bst}
\bibliography{citation.bib}

\newpage
\makeatletter
\renewcommand{\beginappendix}{\appendix
  \section*{\huge\sffamily\bfseries\color{accent-bright} Appendix}
  \addcontentsline{toc}{section}{Appendix}}
\makeatother
\beginappendix
\startcontents[app]
\printcontents[app]{l}{1}{\setcounter{tocdepth}{2}}

\newpage

\section{Related Work}\label{app:related}

\paragraph{Zero-shot mutation-effect prediction.}
Zero-shot predictors rank the effect of mutations without any measurement of the target protein.
Protein language models trained on sequence alone do this: ESM-1v scores a substitution by the masked marginal at its position \citep{meier2021language}, and Tranception adds retrieval of homologous sequences at inference \citep{notin2022tranception}.
ProteinGym collects deep mutational scans to benchmark such models \citep{notin2023proteingym}, and their accuracy depends on how strongly pretraining prefers the wild-type sequence \citep{gordon2025preference}.
ESM-2 \citep{lin2023esm2} is the backbone of \Pev, kept frozen.
Its pretrained language-model head gives the wild-type-marginal log-odds of every edit from one pass over the parent \citep{meier2021language}, which is one of our two zero-shot baselines and also a feature of each candidate in \Pev.
The other is ESM-1v.
\Pev keeps that single pass over the parent and adds heads, trained on measured outcomes, that return calibrated probabilities for each edit on offer.

\paragraph{Machine-learning-guided directed evolution.}
Directed evolution improves a protein through repeated rounds of mutation and selection, and machine learning now guides which variants to assay next.
The classic scheme uses random mutation and artificial selection \citep{romero2009exploring}, and surrogate models fitted to a family's own measurements guide the next round \citep{yang2019mlde}.
Gaussian-process models use their uncertainty to choose the sequences to test next \citep{romero2013gp}, and adaptive machine learning selects sequences for measurement across rounds of training and optimization \citep{hie2022adaptive}.
Machine-learning-assisted evolution works with combinatorial libraries \citep{wu2019machine}, and zero-shot predictors can design its training sets \citep{wittmann2021informed}.
Low-$N$ engineering starts from a handful of characterized variants \citep{biswas2021low}, and few-shot active learning runs on language-model embeddings with a regression head \citep{jiang2025evolvepro}.
\citet{biswas2021low} and \citet{jiang2025evolvepro} encode each candidate's whole sequence, a design that the per-mutant encoder of \cref{sec:frontier} also uses.
Active learning with uncertainty quantification pairs with batch Bayesian optimization \citep{yang2025alde}.
Combining evolutionary density features with assay-labeled data helps when labeled data are limited \citep{hsu2022learning}, and ProteinNPT targets label-scarce, multi-property prediction \citep{notin2023proteinnpt}.
General protein language models have also guided antibody affinity maturation over two rounds \citep{hie2024antibodies}.
Most of these methods fit a model to measurements of the target family.
\Pev is trained once across families and proposes a first round with no measurement of the target.
In the closed loop of \cref{sec:closedloop} it stays ahead of a ridge surrogate and a Gaussian process, both refit on the target's own assays, at every budget.

\paragraph{Fine-tuning and preference alignment.}
Fine-tuning, preference optimization and reinforcement learning adapt models to experimental or energy-based objectives.
Supervised fine-tuning of protein language models improves predictions across tasks and is especially useful for small fitness datasets \citep{schmirler2024finetuning}, and Genome-Factory packages fine-tuning for genomic foundation models \citep{wu2025genomefactory}.
Direct preference optimization trains a language model from preference data in a single stage, without fitting a separate reward model \citep{rafailov2023dpo}.
ProteinDPO uses it to align a structure-conditioned protein language model with experimental stability measurements \citep{widatalla2026proteindpo}.
AlignAb extends antibody direct preference optimization to Pareto optimality over several energy-based objectives for antibody sequence--structure co-design \citep{wen2025alignab}.
Reinforcement learning has also been applied to biological sequence design, with a policy trained on a model fitted to earlier experimental rounds \citep{angermueller2020rl}.
POLO optimizes a lead compound over several turns under a fixed oracle budget, a setting like our closed loop but on molecules \citep{wang2026polo}.
\Pev instead leaves the backbone frozen and trains about 1.6 million parameters of embeddings and heads directly on measured outcomes of candidate edits.
It acts in one pass over a closed set of edits and returns two calibrated probabilities for each edit, one that it is the best action on offer and one that it improves the parent.

\paragraph{Context and coverage in biological foundation models.}
Work on biological foundation models asks how well they carry over to contexts absent from training.
For virtual-cell models, generalization to held-out contexts depends on the diversity of contexts in the training data and not on scale alone \citep{dibaeinia2026virtualcells}.
Cell-JEPA predicts cell embeddings from partial observations instead of reconstructing sparse, noisy counts \citep{elsheikh2026celljepa}.
\Pev applies that view to proteins: one policy trained on many families across peptides, domains and antibodies, and scored on families it never saw.
It predicts a distribution over an action set where a regression head would predict a noisy outcome, and \cref{sec:ablations} compares the two formulations.

\paragraph{Landscapes for closed-loop evaluation.}
Our closed loop runs on a held-out measured landscape.
It is the four-site GB1 landscape \citep{wu2016adaptation}, which also provides low-resource and extrapolative tasks in the FLIP benchmark \citep{dallago2021flip}.
GB1 is reserved from training, so every proposal in the campaign is made on a target outside the training data.

\paragraph{Calibration and proper scoring.}
Calibration asks whether a predicted probability matches how often the event happens.
Strictly proper scoring rules reward honest probabilities \citep{gneiting2007strictly}, and temperature scaling, the single-parameter variant of Platt scaling, was studied for modern networks by \citet{guo2017calibration}.
Uncertainty estimates for protein engineering have been benchmarked for accuracy and calibration \citep{greenman2025uncertainty}.
\Pev trains both heads with strictly proper scoring rules.
It then calibrates them post hoc, with a temperature on the choice logits and an affine map on the improvement logit, both fitted on separate calibration families.
On held-out families, \cref{sec:calibration} shows that this roughly halves the expected calibration error and changes no decision.

\paragraph{System One models.}
System One models make fast, structured decisions that software can use directly.
\citet{typesafe2026jev} describe them as taking structured program state as input and returning type-safe values that carry calibrated probabilities.
Every output comes from one parallel query, with no string generation.
\Pev adopts that interface for protein evolution, with a closed and typed action set.
It trains the interface by supervised minimization of strictly proper scoring rules, with calibration fitted on separate calibration families.


\section{Data, Splits and Licenses}\label{app:data}
This appendix lists the public data sources and their licenses, explains how families are split between training, calibration and evaluation, and names the lineages held out from training.

\begin{table}[H]
  \centering
  \caption{\textbf{Public data sources.} Each source with what it contributes and its license or terms where the source states them; GB1 is reserved for the closed loop.}
  \label{tab:sources}
  \tablestyle\renewcommand{\arraystretch}{1.12}
  \begin{tabularx}{\linewidth}{l Y l}
    \toprule
    \headrow \hd{Source} & \hd{Contributes} & \hd{License or terms}\\
    \midrule
    MegaScale & domain stability & CC BY 4.0\\
    FireProtDB & stability & see primary source\\
    ProteinGym & deep mutational scans & MIT\\
    SKEMPI 2.0 & binding affinity ($\mathrm{p}K_d$) & see primary source\\
    SLiM DMS & peptide binding & CC BY 4.0\\
    AlphaSeq (Ab-14 releases) & antibody binding & CC BY-NC-SA 4.0\\
    AbAgym & antibody--antigen scans & non-commercial\\
    FLAb & antibody mutation series & per-study terms\\
    GB1 four-site & closed-loop landscape only & CC BY 4.0\\
    \bottomrule
  \end{tabularx}
\end{table}

\paragraph{Sources.}
\Pev uses \num{data/public} public sources, listed with their licenses in \cref{tab:sources}.
Of these, \num{data/train_public} supply training and evaluation data, and the GB1 landscape is reserved for the closed loop.
The repository names each source, with its license or terms where the source gives them, and points to the primary data.
The evaluation split behind the benchmark is distributed as one archive whose variant table is pinned by SHA-256, and it carries the terms of the sources from which it is derived.

\paragraph{Splits.}
Splits are made by family, and a family held out for one property stays held out for every property.
The evaluation split holds \num{data/eval_fam} families, \num{data/eval_variants} variants and \num{data/eval_ctx} assay contexts, and the frozen benchmark of \cref{sec:discovery} is drawn from it (App.~\ref{app:defs}).
Every family outside the evaluation split supplies data for training, validation or calibration, and in that role alone: training fits the heads, validation selects the recipe, and the calibration families fit the maps of \cref{sec:calibfit} and appear in no table of results.
Training draws 10{,}000 episodes in each of three epochs and validation 2{,}500, and a calibration pool is accepted at a minimum of three families and 200 episodes (App.~\ref{app:train}).

\paragraph{Held-out lineages.}
Whole lineages are held out from training alongside whole families.
GB1 with every Protein G homologue is held out, matched at the homologue level because MegaScale and ProteinGym carry it under other names, and it serves as the closed-loop landscape.
Others include the Ab-14 lineage in both AlphaSeq releases, the PSD95 PDZ3 domain and twenty MegaScale domains.
The Ab-14 lineage and the PSD95 PDZ3 domain give antibody and domain-binding panels, and the MegaScale domains give domain-stability panels.

\section{Formal Definitions}\label{app:defs}

\paragraph{Beneficial edit.}
An edit is beneficial when its gain exceeds both the measurement noise and a predefined effect floor\longonly{ (\cref{eq:beneficial})}.
The noise $\sigma_a$ combines the replicate standard deviations of the edited sequence and the reference.
The reference value is the median of its replicate measurements, and its standard deviation is the larger of its recorded replicate standard deviation and the standard error across those measurements.
Missing replicate standard deviations are imputed using the median within the assay context, or set to zero if none are recorded.
The effect floor $\delta_\rho$ is fixed before model evaluation at 0.5\,kcal/mol for $\Delta\Delta G$, 1\,$^{\circ}$C for $T_m$, and 0.3 $\mathrm{p}K_d$ units for affinity.
For binding signals, it is half the robust spread of the panel's gains, $0.5\cdot1.4826\cdot\mathrm{MAD}(u)$, where MAD denotes the median absolute deviation.

\paragraph{Benchmark panels.}
The discovery benchmark comprises a fixed list of \num{disc/panels} panels spanning \num{disc/fam} families and \num{disc/cand} measured candidates, released with the code.
Each panel contains a measured unmutated reference and its measured single substitutions within one assay context and protein family.
A panel must contain at least \num{eval/min_cand} candidates with edits inside the encoder's 1{,}022-residue window and at least one beneficial edit.
Panels without beneficial edits are excluded from the discovery benchmark.

\paragraph{Training targets.}
Both heads use soft targets that account for measurement noise.
The choice target $p^\ast$ assigns equal probability to actions in the noise-tie set $\mathcal T$\longonly{ (\cref{eq:target})}.
The improvement target is $y^\ast_a=\Phi(u_a/\sigma_a)$ when $\sigma_a>0$ and $1[u_a>0]$ otherwise, where $\Phi$ is the standard normal cumulative distribution function.
Under Gaussian measurement noise and a flat prior on the gain, this target is the posterior probability that the true gain is positive.
Each episode's loss is weighted by the separation between its two highest-utility actions:
\[
  w=0.1+0.9\,\Phi\!\left(\frac{u_{(1)}-u_{(2)}}{\sqrt{\sigma_{(1)}^2+\sigma_{(2)}^2}}\right),
\]
where $(1)$ and $(2)$ index those actions, including \STOP with $u=0$ and $\sigma=0$.
When the combined noise is positive, the weight is $0.55$ for tied utilities and approaches 1 as their gap grows relative to the noise.
Without recorded noise, the weight is 1.
Calibration instead uses targets based directly on measured outcomes.
The choice target assigns equal probability to the highest-gain edits when their gain is positive, or all probability to \STOP otherwise.
The improvement target is $1[u_a>0]$ for each edit.

\paragraph{Statements and proofs.}
Two propositions state what each head estimates and what parent reuse costs.
{\tcbset{pevthm/.append style={unbreakable}}%
\begin{proposition}[What each head estimates]\label{prop:proper}
Let $\omega=(x,c,\calO)$, let $p^\ast\in\Delta(\calO\cup\{\STOP\})$ be the choice target and $y^\ast_a\in[0,1]$ the improvement target, and let $w\in[0.55,1]$ be the episode weight defined under Training targets.
Write $\E_w[\,\cdot\mid\omega]=\E[w\,\cdot\mid\omega]/\E[w\mid\omega]$ for the expectation under the training distribution reweighted by $w$.
Over all measurable heads, the weighted cross-entropy $\E[-w\sum_j p^\ast_j\log\pchoice[j](\omega)]$ is minimized uniquely by $\pchoice[j](\omega)=\E_w[p^\ast_j\mid\omega]$, and the weighted binary cross-entropy of $\pimp[a](\omega)$ against $y^\ast_a$ by $\pimp[a](\omega)=\E_w[y^\ast_a\mid\omega]$.
With $w\equiv1$ and hard targets, $\pchoice[j](\omega)$ is the posterior probability that $j$ is the best action on offer, ties shared equally, and $\pimp[a](\omega)$ is the posterior probability that the gain of $a$ is positive.
With soft targets they are the weighted conditional means of those targets.
\end{proposition}}
\begin{proof}
Fix $\omega$ and write $q=\pchoice(\omega)$.
The weighted cross-entropy is linear in $w\,p^\ast$, so its conditional mean is $-\sum_j\E[w\,p^\ast_j\mid\omega]\log q_j=\E[w\mid\omega]\,\big(H(\bar p)+\mathrm{KL}(\bar p\,\|\,q)\big)$ with $\bar p=\E_w[p^\ast\mid\omega]\in\Delta(\calO\cup\{\STOP\})$ and $\E[w\mid\omega]>0$.
The KL term is non-negative and vanishes only at $q=\bar p$, which is strict properness \citep{gneiting2007strictly}.
The binary term is the same computation with two outcomes.
\end{proof}

\begin{proposition}[Cost of parent reuse]\label{prop:cost}
Let $C_{\mathrm{enc}}$ be the cost of one backbone pass over a length-$L$ sequence and $C_{\mathrm{head}}$ the cost of scoring one candidate from its features.
For a decision over $N$ candidates, parent reuse costs $C_{\mathrm{reuse}}=C_{\mathrm{enc}}+N\,C_{\mathrm{head}}$.
Encoding every mutant separately costs $C_{\mathrm{pm}}=(N+1)\,C_{\mathrm{enc}}+N\,C_{\mathrm{head}}$, because each mutant has the parent's length and the parent is still encoded for the \STOP head and the language-model features.
The ratio $S(N)=C_{\mathrm{pm}}/C_{\mathrm{reuse}}$ is increasing in $N$ and tends to $1+C_{\mathrm{enc}}/C_{\mathrm{head}}$.
\end{proposition}
\begin{proof}
$S(N)=\dfrac{N(C_{\mathrm{enc}}+C_{\mathrm{head}})+C_{\mathrm{enc}}}{N\,C_{\mathrm{head}}+C_{\mathrm{enc}}}=1+\dfrac{C_{\mathrm{enc}}}{C_{\mathrm{head}}+C_{\mathrm{enc}}/N}$, which increases in $N$ and tends to $1+C_{\mathrm{enc}}/C_{\mathrm{head}}$ from below as $N\to\infty$.
\end{proof}

\paragraph{Screen metrics.}
The screen scores each panel on the ten edits it ranks first and on its full ranking.
Hit rate@$k$ averages the labels of the top $k$ edits of a panel, with $k=10$ since every panel holds at least \num{eval/min_cand} candidates.
Enrichment@$k$ divides it by the panel prevalence, so random ranking scores one in expectation.
AUROC is the Mann--Whitney statistic per panel, and wells to first hit is the one-based rank of the first beneficial edit in the full ranking.
Perfect ranking places the beneficial edits first, ordered by measured effect, and the random baseline averages twenty permutations per panel.
\Pev ranks each panel by its choice logits, for both objectives.
Zero-shot ESM-2 ranks by the wild-type-marginal log-odds of the edit from one pass over the parent, and ESM-1v by the masked marginal at the edited position, averaged over its five published checkpoints.
The screen's tables are means over panels.

\paragraph{Transfer metrics.}
The transfer diagnostic labels an edit by the noise margin alone and scores one decision per panel.
An edit clears its noise there when $u_a>\sigma_a$, with $\sigma_a$ the replicate sd of the difference ($u_a>0$ where none is recorded) and no effect floor.
Success@1 is one if the chosen edit clears its noise, or if \STOP is chosen when no edit does.
\Pev and its variants can choose \STOP, while zero-shot ESM-2 and random always choose an edit.
Panel AUROC uses the same label and does not depend on \STOP.
Perfect decisions choose an edit that clears its noise when one exists and \STOP otherwise, and rank the edits that clear their noise first, so both metrics equal one.
Rows are weighted by class, then objective, then (family, context), then panel.

\paragraph{Closed-loop and frontier metrics.}
The closed loop scores the best sequence a campaign reaches, and the frontier scores quality against speed.
The best observed fitness after $B$ assays is $F_B=\max_{t\le B}y(x_t)$, where $x_t$ is the $t$-th assayed sequence and fitness is relative to the wild type, $y(\mathrm{WT})=1$.
\longonly{The perfect one-step ranking reference follows the same protocol but assays, each round, the \num{gb/batch} unassayed neighbors of the current parent with the highest measured fitness.
The share of the achievable gap in subfigure (b) of \cref{fig:gb1} is $(F_B-F_B^{\mathrm{rnd}})/(F_B^{\mathrm{perfect}}-F_B^{\mathrm{rnd}})$, with $F_B^{\mathrm{rnd}}$ the value for random selection.}
For a panel and the set $R$ of the top eight edits returned, $P=\max(0,\max_{a\in R}u_a)/\max_{a\in\calO}u_a$, with $P=0$ when no edit improves; it is averaged within each class, family and context and then over the three classes with equal weight.
Throughput is the reciprocal of the median seconds per candidate over repeated timed requests of 1{,}000 candidates.

\paragraph{Calibration error.}
Expected calibration error is the weighted gap between predicted probability and observed frequency.
It is computed on the held-out test families, on panels that each hold a measured parent with its measured single-edit neighbors, and never on the calibration families on which the maps are fitted.
It uses ten equal-width bins $m$, the last including 1, and $\mathrm{ECE}=\sum_m \tfrac{W_m}{W}\,\lvert\mathrm{pred}_m-\mathrm{freq}_m\rvert$, where $W_m$ is the total weight of the forecasts in bin $m$, $W=\sum_m W_m$, and $\mathrm{pred}_m$ and $\mathrm{freq}_m$ are their weighted mean predicted probability and weighted observed frequency.
Weights give equal mass to each objective, then to each (family, context) within it, then to each panel, with candidates equal within a panel.
The improvement forecast is scored against $1[u_a>0]$ for every candidate.
The choice forecast is the probability of the top-ranked edit, scored against the hard choice target of Training targets, so it counts as correct when that edit is a best positive-gain edit of its panel.

\paragraph{Uncertainty.}
Intervals come from a paired bootstrap over families on the screen and over initializations in the closed loop.
One family can supply several panels, so the screen resamples the per-family means of the two methods compared 2{,}000 times, and the intervals are 95\% percentile intervals.

\section{Training and Evaluation Details}\label{app:train}

\paragraph{Model.}
The frozen ESM-2 650M backbone has hidden size $d=1280$, and the trained parts hold 1{,}560{,}643 parameters.
The context embedding $q_c$ has $32+32+2\times129=322$ dimensions: a 32-dimensional objective embedding, a 32-dimensional readout embedding, and the mean-pooled target and partner states, each projected from 1280 to 128 dimensions with a presence flag.
With a 32-dimensional residue embedding shared by $e_r$ and $e_{x_i}$, the candidate feature $\phi_a=\big[h_i;\ \bar h;\ e_r;\ q_c;\ e_{x_i};\ \ell_a\big]$ has $1280+1280+32+322+32+2=2{,}948$ dimensions.
The candidate head maps it through two hidden layers of width 256 (GELU, dropout 0.1) to the choice and improvement logits.
The \STOP head maps $[\bar h;q_c]$, of $1280+322=1{,}602$ dimensions, through one hidden layer of width 256 to the \STOP logit.

\paragraph{Optimization.}
\Pev is trained with AdamW for three epochs of 10{,}000 wild-type-scan episodes.
The learning rate is $3\times10^{-4}$, weight decay 0.01, gradient clipping 1.0, and training uses bf16 autocast and gradient accumulation over 8 episodes.
Episodes are drawn evenly across the sequence-class and objective cells, and validation loss is computed on 2{,}500 validation episodes.

\paragraph{Calibration fit.}
Calibration is fitted separately for each objective on the calibration families.
L-BFGS starts from the identity and fits $T$ on the weighted choice log-loss and $(\alpha,\beta)$ on the weighted binary cross-entropy, with equal mass per objective, then (family, context), then episode.
The bounds are $\log T,\log\alpha\in[-5,5]$ and $\beta\in[-10,10]$.
Each map is fitted at the first of four pools that gives an accepted fit: the objective's own calibration episodes from its primary sources, all calibration episodes of that objective, both objectives pooled, and the identity.
A fit is accepted when its pool has at least three families and 200 episodes and, for the affine map, both outcome classes, and when it does not run to a flattening bound.
The pool reached is published with the parameters.

\paragraph{Closed-loop rollout.}
Every arm runs the rollout of Algorithm~2 and spends its whole budget of \num{gb/budget} assays in batches of \num{gb/batch}, counting the initial set.
Each of the \num{gb/n_init} initializations is a measured sequence and seven of its neighbors, chosen by hash without reading any outcome.
\Pev scores by calibrated $\pimp$, the ridge surrogate by its predicted fitness, the Gaussian process by expected improvement, and variants without a trained improvement head by their own score.
The ridge surrogate uses one-hot features over the four sites ($4\times20$) with penalty 1 and is refit in closed form every round on all measured data.
The Gaussian process uses the same features with a squared-exponential kernel, refit the same way.
Neither comparator chooses \STOP.

\paragraph{Throughput.}
Throughput is measured on one GPU.
Each request holds single substitutions of one held-out parent, and timing is CUDA-synchronized with the frozen context cache warmed first.
\note{Fill in the GPU model used for the throughput sweep (probably the NVIDIA GeForce RTX 3090 Ti, 24\,GB; owner to confirm).}

\paragraph{Reproducibility.}
Two commands reproduce the discovery tables from the released checkpoint and the evaluation archive.
\Mode{make~check} confirms the variant table, the evaluation split and the \num{disc/panels} panels against their pins, and \Mode{make~discovery} writes every discovery table.

\paragraph{Broader impact.}
\Pev edits existing sequences toward measured assay outcomes.
It chooses among single substitutions of a given sequence rather than generating new ones, and it is trained on public data.
We release a checkpoint, a frozen evaluation split and the code of the discovery screen.

\section{Algorithms}\label{app:algs}
Algorithm 1 gives one \Pev decision, and Algorithm 2 wraps it in a closed-loop campaign.

\begin{algobox}{Algorithm 1: one \Pev decision}
\begin{enumerate}[label=\arabic*:,itemsep=1pt]
  \item \textbf{Input:} parent $x$, context $c$, offered edits $\calO$ (the measured substitutions in a screen, the unassayed ones in a campaign); calibration $(T,\alpha,\beta)$ for the objective of $c$.
  \item Encode the parent once with the frozen backbone, $h\leftarrow\mathrm{Enc}(x)$, and embed the context as $q_c$.
  \item For every edit $a=(i,r)\in\calO$ build $\phi_a$ from $h_i$, $\bar h$, $e_r$, $q_c$, $e_{x_i}$ and $\ell_a$, and compute $(z_a,v_a)$ in one batched call.
  \item Compute the \STOP logit $z_{\STOP}$ from $[\bar h;q_c]$ with the \STOP head.
  \item $\pchoice\leftarrow\softmax\big(([z_a]_a,\,z_{\STOP})/T\big)$ and $\pimp[a]\leftarrow\sigma(\alpha v_a+\beta)$.
  \item \textbf{Return} $\hat a=\argmax\pchoice$ (an edit or \STOP), with $\pchoice$ and $\pimp$.
\end{enumerate}
\end{algobox}

\begin{algobox}{Algorithm 2: closed-loop campaign with \Pev}
\begin{enumerate}[label=\arabic*:,itemsep=1pt]
  \item \textbf{Input:} assayed starting set $S$, budget $B$, batch size $b$, the decision rule of Algorithm 1.
  \item \textbf{While} $|S|<B$:
  \item \quad parent $\leftarrow$ the best assayed sequence in $S$ that is not retired and has unassayed neighbors.
  \item \quad Score every unassayed candidate edit of the parent with Algorithm 1.
  \item \quad Rank by $\pimp$, take the best edit at each site, then fill the batch of $b$ by rank.
  \item \quad Assay the batch and add it to $S$; if Algorithm 1 returned $\hat a=\STOP$, retire the parent.
  \item \textbf{Return} the best observed fitness in $S$.
\end{enumerate}
\end{algobox}


\section{LLM Usage}\label{app:llm}
Coding agents helped implement the model, the training pipeline and the evaluation benchmark, and helped edit this paper.
The authors proposed the decision formulation, designed the experiments, made the calibration decisions, and checked every claim and every number reported here against the benchmark outputs.

\end{document}